# Game theory and scholarly publishing: premises for an agreement around open access

Abdelghani Maddi[1]

[1] *abdelghani.maddi@hceres.fr*
Observatoire des Sciences et Techniques, Hcéres, 2 Rue Albert Einstein, Paris, 75013 France.

## Abstract

Stakeholders in research and scientific publishing are gradually joining the Open-Access (OA) movement, which is gaining momentum to become nowadays at the heart of scientific policies in high-income countries. The rise of OA generates profound changes in the chain of production and dissemination of knowledge. Free access to peer-reviewed research methods and results has contributed to the dynamics of science observed in recent years. The modes of publication and access have also evolved; the classic model, based on journal subscriptions is gradually giving way to new economic models that have appeared with the arrival of OA.

The objective of this article is twofold. First, propose a model for the publishing market based on the literature as well as on changes in open science policies. Second, analyze publishing strategies of publishers and institutions. To do so, we relied on game theory in economics. The analysis carried out underlines the complexity of the scholarly publishing market. In pure strategies, three equilibria are possible: the two players converge towards an OA model; publishers opt for a hybrid model while institutions for an OA model, and finally publishers take the initiative of publication in OA while institutions remain on a traditional model. In mixed strategies, in an uncertain environment, the results show that there is one and only one equilibrium: the OA model.

## Acknowledgment

The author would like to thank Luis Miotti for his careful reading and his precious remarks, which contributed to improve the paper in a significant way. The author would also like to thank the French "*Observatoire des Sciences et Techniques*" for funding the article processing charges.

---

[1] Corresponding author : Abdelghani Maddi, Observatoire des Sciences et Techniques, Hcéres, 2 Rue Albert Einstein, Paris, 75013, France, T. 33 (0)1 55 55 61 48.

# Introduction

The arrival of digital technology at the start of the 21$^{st}$ century has completely changed the world of scientific publishing [1–4]. Publishers' business strategies have seen several innovations regarding the pricing of research products [4]. In the traditional model of scientific publishing, also called the "reader-pays" economic model, readers access to publications only through subscriptions to scientific journals (the journal's main source of income) [5]. One of the only strategies available to publishers in this business model is to adjust the price according to the level of demand in the market which depends greatly on the quality of the journals, in addition to the number of publications received [6,7].

With the advent of Open Access (OA), other business models have emerged. The best known are the "author-pays" and "hybrid" economic models. In the first one, the research results (publications, data, etc.) are freely accessible to everyone without any geographic or time limit [8,9]. In this model, it is the authors who pay (usually through their institution or funder) publication fees (*Article Processing Charges – APCs*) to make their research freely accessible [10,11]. In the hybrid model there is a "cohabitation" of the two models "reader-pays" and "author-pays" [7,12]. The hybrid model has been widely adopted from 2013 by the world's five largest publishers: Elsevier, Springer, Taylor and Francis, Sage and John Wiley and Sons, which account for over 70% of the journal publishing market [12]. Thus, in the same journal, there could be OA publications for which the authors have paid fees and publications accessible by subscription. It should be emphasized here that, in the hybrid model, if an institution, involved in the OA movement, subscribes to this journal, it systematically pays twice for its OA articles published there (APCs and subscription).

There are other economic models, such as the "sponsor-pays" model when a journal is fully funded by an organization or an association [13–15]. In this model, neither the authors nor the readers pay to publish or read the articles. There is also the "freemium" model in which the publisher makes all or part of a publication available in a simple format (html or text for example), then remunerating himself on access to more convenient formats (e.g. pdf) for the reader, or on access to additional information, for a variable price [16–18].

The proliferation of business models for scientific publishing is a real boon for publishers [19–21]. Björk (2017) has shown that, on the one hand, large publishers continue to dominate the market, which operates under the rules of an oligopolistic market, and on the other hand, the profits made by the latter have increased considerably with the arrival of the OA. The situation is different for institutions and funders. In addition to the costs of subscriptions that must bear to guarantee their researchers access to publications, they are now increasingly led to pay OA publication costs (APCs) which can reach 5,000 euros for a single publication. An increasing number of stakeholders in scholarly publishing consider that the current system of subscriptions is becoming anachronistic and it is imperative to go to fully OA model. The concept of the "Big Deal" then appeared to designate licensing agreements between publishers and institutions (or funders) including both the price of APCs and subscriptions [22–25].

Beyond purely economic considerations, the scientific world is witnessing the rise of a movement in favor of OA which consists in saying that insofar as research is mainly financed by public funds, the results of research should be too [26–29]. In addition, moving to a model entirely in OA allows better equity, in particular for researchers whose institutions do not have sufficient means to finance both subscriptions and APCs [30–33].

A new paradigm, around the "Big deal", in the scientific publishing market was then born to allow better dissemination of research products [34]. This is about empowering researchers not only to have free access to the results of peer-reviewed research, but also to publish "freely" without paying supplementary fees. Within the scientific community, we commonly speak of the "Read and Publish" and "Publish and Read" license agreements [35–38]. The two concepts are close, with the difference that the first one places more emphasis on being able to "read / access" freely, while in the second one the emphasis is on the ability to publish in OA without paying APCs.

Although the question of the impact of digital transformations on the publishing market is largely invested in the literature, little work analyzes the question from an economic point of view (e.g. market equilibrium). Much of the current research on the publishing market reports on the various existing models and their development [34,39–44], characteristics of the competition [4,21,45,46], costs of scientific publishing in the OA era [7,20,47,48], self-archiving practices for research results [2,49], the difficulties to switch to an OA model and the negotiation power of publishers, or the future challenges of scientific publishing [11,19,21,50–53], including in emerging countries [30,31,33,54–56].

The purpose of this article is to examine the issue through the prism of economic market analysis using game theory. First, the objective is to model the publishing market objectively using the results of the literature and the dynamics observed in recent years within both scientific communities and public policies. Second, study the possible equilibria to which the market can converge using the principle of "best response" in game theory and the Nash equilibrium. We also analyzed the stability of equilibria and their Pareto optimality.

It is important to clarify that in the present study it is mainly about the so-called "gold" or "diamond" roads of OA. That is, the "author-pays" and "sponsor-pays" business models. The aim is to analyze markets where OA is the result of an institutional effort. The other models: "green" or "bronze" road are dissemination models that reflect the efforts of researchers and journals to make certain publications accessible (self-archiving).

<mark>Plan</mark>

## Literature review: the scholarly publishing market in game theory

Although not numerous, the existing studies that use game theory to analyze the publishing market provide a sufficiently developed theoretical framework. The publishing market defined as the meeting place of three types of economic agents: publishers, authors and readers.

In accordance with the axioms of rationality in behavioral microeconomics, each of these agents implements strategies to maximize its utility. For the authors it is a question of maximizing their "reputation", source of funding, which depends on both the academic and societal impact, for the readers the maximum utility lies in the appropriate choice of articles to read according to their intrinsic (and not perceived) quality, while for publishers it is about maximizing profits, given the strategies adopted by both authors and readers [12,34,57,58].

[57] have built a theoretical model from the strategies ($S_i$) of authors who make a trade-off between "publishing in open access" (O) and "not publishing in open access" (Ø). [57] compare the payoff matrices of the authors in a classical and quantum game by approaching three types of game: a zero-sum game, the "prisoners' dilemma" and a "stag hunt" version of the game. In their model, the payoff is represented by the reputation obtained as a result of the publication. For two researchers A and B the game tree is as follows:

**Fig. 1. Classical tree of the open access game.**

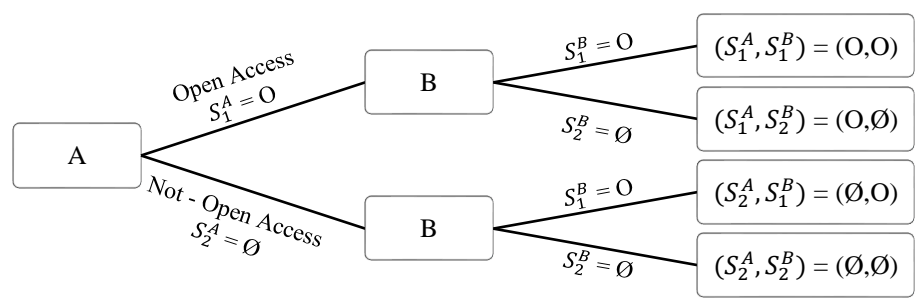

Source: Hanauske, Bernius and Dugall (2007), p. 651.

The global strategic space $\mathscr{S}$ is the Cartesian product of the two strategies of A and B:

$$\mathscr{S} = \mathscr{S}^A \times \mathscr{S}^B = \{(O,O), (O, Ø), (Ø,O), (Ø, Ø)\}$$

The payoff matrix for this game is as follow:

**Table 1: General open access payoff matrix**

| A/B | O | Ø |
|---|---|---|

| O | $(r+\delta, r+\delta)$ | $(r-\alpha, r+\beta)$ |
|---|---|---|
| Ø | $(r+\beta, r-\alpha)$ | $(r, r)$ |

Source: Hanauske, Bernius and Dugall (2007), p. 652.

The parameters $\alpha \, et \, \beta$ ($\alpha, \beta \geq 0$) respectively represent the decrease and the increase in reputation depending on the chosen strategy. $\delta$ represents the advantage that A and B derive if they simultaneously choose to publish in Open Access.

In this game, the reasoning is as follows: in a traditional (non-OA) model, the reputation of researchers' work largely depends on the journals in which they publish. In the context where the study of Hanauske, Bernius and Dugall (2007) is carried out (ie mid-2000s), the major journals in most disciplines have not yet adopted an OA model, or at least a hybrid model. The great transformation took place a few years later (in 2013) as underlined above. Therefore, publishing OA for an individual researcher was a risk as well-established journals in the market are subscription-based. However, publishing in OA allows better visibility of research and therefore better impact and increased reputation. Hanauske, Bernius and Dugall (2007) have shown that a strategy in favor of OA cannot be chosen at the time the study is carried out, due to a lack of sufficient incentive for researchers to publish freely access (especially from the point of view of the research evaluation where the attention is more focused on impact than on openness). The high level of risk made the researchers stuck in a Nash equilibrium with strategies (Ø, Ø). The authors concluded that high pressure from the scientific community to publish in OA could change the characteristics of the publishing market and make it more open, as is the case in certain disciplines such as mathematics or physics.

The study by [34] came to similar conclusions regarding the lack of incentive for researchers to publish in OA, which slows the process of paradigm shift towards a more open model. Therefore, the emphasis on the impact of publications is so strong that researchers are more or less forced to choose a journal based on its impact factor and not its openness status. In addition, the authors underline the complexity of the landscape on the side of the open access publication with a multitude of approaches and models, which constitutes a brake on the development of the open access model.

[59] provided an improved version of the model of Hanauske, Bernius and Dugall (2007) who develop a two-researcher approach with symmetrical strategies and gains. In the new version proposed by (Habermann and Habermann, 2009) it is rather an asymmetrical game with a conflict of interest between researchers on the one hand and publishers on the other.

Thus, in this configuration, there are two payoff matrices; one for publishers and one for authors. The two players have two strategies: to publish in OA (one notes respectively for the authors and the editors: $(s_1, p_1)$ or to choose the traditional way $(s_2, p_2)$. R> 0 represents the gain (reputation) of the authors, 0 <r <R is the decrease in the author's reputation if he chooses the $s_1$ strategy (for the same reasons mentioned in Hanauske, Bernius and Dugall (2007)). (Habermann and Habermann, 2009) integrate the impact of publications (I) as an additional source of gain for both authors and publishers.

This impact drops by $0 < \tau < I$ in a traditional model (less visibility). In terms of expenditure, Habermann and Habermann (2009) integrated three types: L>0 corresponds to expenditure in an OA model. According to the authors, L is a cost borne equally by the authors and the publishers $(s_1, p_1)$. G>0 represents the price of subscriptions and / or APC paid by researchers. Finally, the authors also integrate P> 0 to denote the high profit of publishers from very expensive journals in a traditional model.

**Table 3: Payoffs of authors and publishers according to strategies**

| Strategies | Author's payoff | Publishers payoff |
| --- | --- | --- |
| $s_1 \leftrightarrow p_1$ | $(R - r) + I - L/2 - G$ | $G + I - L/2$ |
| $s_1 \leftrightarrow p_2$ | $(R - r) + I - L$ | 0 |
| $s_2 \leftrightarrow p_1$ | $R + (I - \tau) - G$ | $G + (I - \tau) - L$ |
| $s_2 \leftrightarrow p_2$ | $R + (I - \tau) - G - P$ | $G + (I - \tau) + P$ |

Hanauske, Bernius and Dugall (2007) described an unstable equilibrium with an oscillation in the form of a circle between several strategies:

**Fig. 2: Iterative change in strategies between authors and publishers**

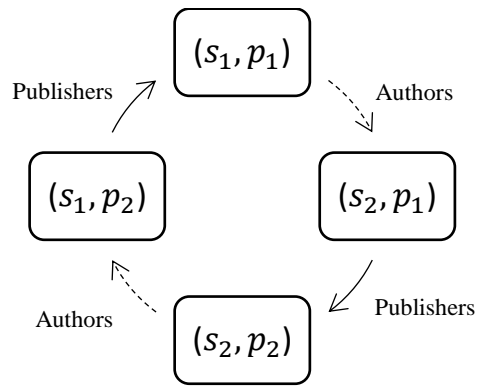

If authors and publishers simultaneously adopt an OA strategy $(s_1, p_1)$, authors can increase their gain by changing their strategy from $s_1$ to $s_2$ since $R + (I - \tau) - G > (R - r) + I - L/2 - G$. The game therefore switches to a strategy $(s_2, p_1)$. This new configuration also encourages the editors to modify their strategy to go to $p_2$. This will allow them to increase their gains $G + (I - \tau) + P > G + (I - \tau) - L$. The global strategy then becomes $(s_2, p_2)$.

Authors can improve their gain by switching to a strategy $(s_1, p_2)$ and so on. In this model, equilibrium is therefore a time dependent function.

In the study of [12] the reasoning is completely different. Publishing in OA does not constitute a risk or loss of reputation for authors (as is the case in the previous studies), on the contrary. [12], relied on the bibliometric literature to emphasize that researchers who publish in OA are generally good researchers and that several fundamental journals in different disciplines have adopted the OA or hybrid model from 2013 The authors hypothesized that choosing to publish in OA could serve as a quality signal for the scientific community (readers) and for publishers. (Besancenot and Vranceanu, 2017) therefore built their model using signal theory. The authors make a trade-off between an OA strategy (noted A) and a traditional strategy (T) (not OA). µ is the proportion of high quality papers ($\theta^H$). By deduction, (1- µ) is the number of low / medium quality papers ($\theta^L$). As good articles are less frequent (cf. distribution laws of citations in bibliometrics), µ <1/2.

The gain function, designated by the product between the size of the readership ($\delta_S$) and the quality, of an author of type $i \in \{H, L\}$, for a strategy $S \in \{A, T\}$ is composed of the intrinsic quality $\theta^i$ and the perceived quality $E(\theta|[S])$. $\lambda$ is a structural parameter to measure the weight of the two terms. The lower $\lambda$, the more important the intrinsic quality. The authors assume that $\lambda < 0,5$. Finally, c denotes the price of APCs in an OA (or hybrid) model. Therefore, $I_S$ takes 0 in a traditional model and 1 in an OA model. The gain function is as follow:

$$U^i = \delta_S[(1-\lambda)\theta^i + \lambda E(\theta|[S])] - I_S c$$

For publishers, the payoff is represented by the income R which depends on the frequency of authors who have opted for OA multiplied by the price of APCs: $\varphi c$, plus the frequency of authors who have opted for a traditional model multiplied by income subscriptions $(1 - \varphi)E[\theta|T]$. The editor's gain function is therefore:

$$R = \varphi c + (1 - \varphi)E[\theta|T]$$

Besancenot and Vranceanu (2017) have defined several types of equilibria for publishers in this market depending on the strategies adopted by the authors. An equilibrium with: opposing strategies, identical strategies and hybrid strategies. The table below summarizes the different situations:

**Table 2: Types of equilibria according to the authors' strategies**

| # | Equilibrium type | Publishing market | Function of editors at equilibrium (see Besancenot and Vranceanu | Author' strategies |
|---|---|---|---|---|

| # | | | | |
|---|---|---|---|---|
| | | | (2017), for demonstrations / conditions) | |
| 1 | Separating equilibrium | Global | $R_{Sep}(c) = c\mu + (1 - \mu)\theta^L$ | Opposing strategies: good researchers choose OA and less good traditional. |
| 2 | Pooling equilibrium | Open Access | $R_{P\,A}(c) = c$ | All Researchers Choose OA: Occurs when the cost of publishing in OA is not high. |
| 3 | | Traditional | $R_{P\,T} = E[\theta\|T] = \mu\theta^H + (1 - \mu)\theta^L$ | All researchers choose the traditional model: Occurs when the cost of publishing in OA is high. |
| 4 | Hybrid equilibrium | Hybrid 1 | $R_{H1}(c) = \theta^L - \mu\lambda\delta_A(\theta^H - \theta^L)[\dfrac{(\theta^L - c)}{c - (\delta_A - \delta_T)\theta^L}]$ | All the good researchers choose OA, the less good ones choose both (sometimes OA and sometimes traditional). |
| 5 | | Hybrid 2 | $R_{H2}(c) = \varphi c + (1 + \varphi)E[\theta\|T]$ | All the less good researchers choose traditional, the good ones choose both (sometimes OA and sometimes traditional). |

(Besancenot and Vranceanu, 2017) have shown that the publishers' preference for a particular equilibrium according to the choices of the authors depends on four elements: quality of the journal, the size of the readership in a traditional model (number of subscriptions notably), the difference in terms of accessibility / readership in the two models (OA and traditional) ($\delta_A - \delta_T$) and the quality gap between high impact and low / medium impact researchers. The table below summarizes the publisher preferences:

**Table 3: Journal types and preferred equilibria**

| Journal type | $\delta_T$ | Accessibility gap ($\delta A - \delta T$) | Quality gap ($\theta^H - \theta^L$) | Publisher preferred equilibrium (#) |
|---|---|---|---|---|
| | | | | |

| Leading | High | Low | Low | 1 |
| Specialized (good) | High | High | Low | 1 et 2 |
| Second tier | Low | High | High | 1 et 3 |

Source: Besancenot and Vranceanu (2017), p. 20.

We can see in the table that the separating equilibrium (# 1: all good researchers choose to publish in OA and the less good ones in non-OA) is preferred whatever the situation. On the other hand, in the case of intermediate quality journals (good / medium quality) where the readership gap between OA and traditional is high (in favor of OA obviously) and the quality gap between good and less good researchers is low , the editors will prefer, in addition to # 1, a model in which all the researchers publish in OA (mixing equilibrium, # 2). Finally, for journals with low impact, low readership and gaps in accessibility and high quality, publishers will prefer, in addition to the separator balance, a traditional mixing balance (# 3).

Until then, the various studies have built their reasoning at the finest mesh level: researchers, while OA publication strategies are largely influenced by institutions. At least, they are the ones bearing the costs of subscriptions and APCs. Institutions and funders devote a large budget to the scientific publishing market and open science; the market is estimated at over 5 billion euros. Therefore, in this article, the reasoning is built from the institutions considered the center of decision, in addition to the publishers.

## Modeling the publishing market

**Main assumptions**

As pointed out previously, in the current publishing market, there are three main economic models: subscription based, open access and hybrid that combines the first two. In this section, we rely on the results of the literature to model the level of production and accessibility, academic impact and costs / profits in each of the three economic models.

### General Axioms on the Scholarly Publishing Market

- *Axiom 1: institutions determines the publication strategy in author-pays economic model.*

Authors rarely pay APCs with their means; they use funding from their institutions. Thus, the choice to publish in Gold Open Access strongly depends on the agreement of the institution and its publication policy. However, authors can choose to dump their publications in an open archive after the embargo period. In this case, it is more of a dissemination/diffusion strategy (green OA) than a publication strategy. This is why we are reasoning here from the institutions.

- *Axiom 2: the publishing market is subject to herd behaviors regarding the choice of publication strategies.*

[60] highlighted the mimetic behavior of agents in financial markets and the influence this has on prices. Several studies in other areas go in the same direction to show that individual choices can be driven by the belief that each player makes on the belief of others. In waterfall models, the lack of information generates mimetic behaviors consisting in reproducing the same choices of peers. Thus once a waterfall is triggered, individual decisions no longer depend on the signals perceived in a private capacity by the agents, but rather by collective knowledge and movement [61,62].

By analogy, a movement in favor of open access, for example, from the major publishers (following the pressure exerted by the institutions), can sending a strong signal to the rest of the publishers of a paradigm shift. This can trigger a series of mimetic behaviors to align with new market transformations. This reasoning is also consistent from the point of view of the institutions.

- *Axiom 3: A «convergence coefficient» determines the speed of convergence of the players (economic agents) towards a given publishing model.*

As presented in Axiom 2, in the event of a cascading movement, the entire market could converge into a single future strategy. We represent this as a convex curve that denotes the gradual increase in the share of journals that adopt a given publishing business model. The slope of the curve measure the speed of convergence at each instant. The latter depends on several factors, which may be external to the market (such as State intervention to support a given model - OA for example).

**Fig. 3: Convergence coefficient**

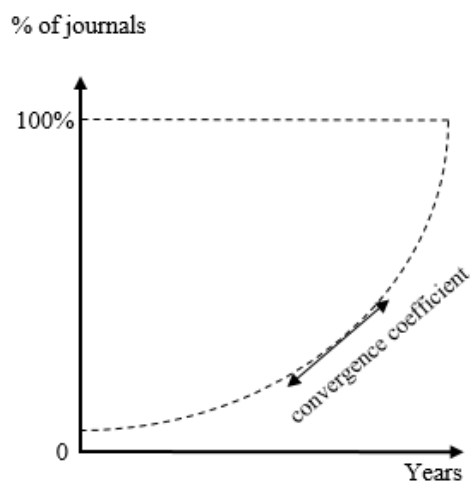

- *Axiom 4: Institutions and publishers are rational and seek to maximize their utility.*

This rationality assumption states that both players seek to maximize their utility when possible. For institutions, this means maximizing the number of publications to which they have access and those they can publish, in terms of their visibility and at the cost incurred. On the publisher's side, utility depends primarily on profit and reputation.

### Production and access

We assume that each model allows producing a number of publications denoted $P^C$, $P^{OA}$ for the subscription and OA models respectively. In terms of dissemination of publications and their accessibility to the scientific community, each of these two models allows a different level of dissemination that we denote $A^C$, $A^{OA}$ respectively. In the context of the hybrid model, the number of publications corresponds to the value $P^H = [\lambda P^{OA} + (1 - \lambda)P^C]$ with $\lambda$ the proportion of open access publications. Likewise, the number of publications accessible in this model is $A^H = [\lambda A^{OA} + (1 - \lambda)A^C]$. Since OA is rather an exception in the hybrid model, one can reasonably assume that $\lambda < 0.5$.

In terms of production, as shown in the literature [5], a transition to an OA model would increase the overall production of scientific publications as well as their dissemination. Likewise, as shown in [7,63], researchers are increasingly inclined to publish in OA even in hybrid journals. Although many institutions and funders advise against publishing OA in hybrid journals [25,64], the market share of hybrid journals in all OA publications has increased dramatically with the OA movement [12,65–67]. It is therefore possible to establish the following relations: $P^{OA} > P^H > P^C$ on the one hand and $A^{OA} > A^H > A^C$ on the other hand.

### Academic and societal impact

There is an abundant literature on academic impact based on the openness status of publications or journals. It is important to distinguish between OA journals on the one hand and OA publications. Overall, open access publications are more visible and therefore on average more cited [68–70]. On the other hand, from a journal perspective, the literature shows that high impact journals are mostly hybrid. The average impact of open access journals is lower, but still higher than that of completely closed (subscription based) journals. The reason behind these results is that journals that are well established in the market are generally hybrids [9,71–73]. These journals therefore receive more citations than fully open access journals, many of which are recently created.

In short, the academic impact depends on both the accessibility of publications (openness) and the notoriety of the journals. We notice: $I^j(A, N)$, with $I$ the journal impact, $j$ the publishing model of journal, "A" the degree of accessibility of the publications and "N" the notoriety of the journal. For the three publishing models we establish: $I^H(A^H, N) \approx I^{OA}(A^{OA}, N) > I^C(A^C, N)$.

### Costs and Profits

From the point of view of publication costs, the hybrid model is the most expensive insofar as for a good part of the publications, the institutions pay both the APCs to publish in OA and the subscriptions to have access to the publications. On the other hand, it can be assumed that in a successful transition to an OA model, the fees that institutions would incur will be lower than the fees paid in a subscription model. The reason why the costs in an OA model would be lower is in the fact that the institutions have an increasing negotiating power, since scientific publications are produced and reviewed by researchers. The added value provided by publishers is decreasing with the rise of self-archiving (and open peer review) of publications with dedicated platforms. The costs in an OA business model are mainly those of APCs, while in a subscription business model they are the subscription prices and finally in a hybrid model the costs are made up of the subscription prices in addition to the total cost of the publications in OA. This cost depends on the number of OA publications in hybrid journals.

We can therefore note: $C^H(\sum \varphi * APC + S) > C^C(S) > C^{OA}(APC)$. With $C$ the publication cost. Conversely, and by deduction, publishers' profits will therefore be higher in a hybrid model and less important in an OA model where institutions and funders carry out negotiations. We therefore note: $B^H < B^C < B^{OA}$ With B the profits of the publishers.

### Interactions of the game and payoffs

The payoffs of institutions and funders depend on the three elements presented in the previous section, namely the number of publications produced and to which they have access, the citation impact and the costs incurred. The gains also depend on the interaction that have institutions with publishers.

On the publishers' side, payoffs depend on the profit they make as well as the reputation of the journals they publish. The reputation of journals is strongly linked to their academic and societal impact, but also the number of contributions they receive. The profit depends on quantity of the articles published multiplied by the price, from which the editorial costs are deducted. Prices depend on the economic model adopted, as does the quantity of publications. Thus, in the case of an OA model, APCs represent the price of publications, whereas in a subscription-based model, it is the average price of a subscription. On the other hand, the costs can be divided into fixed costs and variable costs. Fixed costs are stable regardless of the publication model adopted, while variable costs depend on the number of publications received and published.

In summary, the profit function of a publisher $\lambda$ can be written for an economic model $i$:

$$\pi_{\lambda_i} = \sum P_{ij} * Pe_{ij} - (FC_j + VC_j(P_i))$$

with $P_{ij}$ the number of publications of the journal $j$ in the publishing model $i$, $Pe_{ij}$ the price applied by the journal $j$ which depends on the publishing model $i$. $FC_j$ the fixed costs of the journal $j$, $VC_j$ the variable costs of the journal $j$ which are a function of the number of publications $P_i$. Here, for simplicity, we consider that $P_i$ includes all the articles received including those that are rejected (because they also generate processing costs).

There are two main situations:

**Both players choose the same strategy**

We note the payoffs of the institutions for the models: OA, hybrid and on subscription $U_I^H, U_I^{OA}, U_I^C$ respectively. Likewise, publishers' payoffs for each model are noted $U_\lambda^H, U_\lambda^{OA}, U_\lambda^C$.

We can distinguish three possible situations

- ***Both choose an OA model***

The formalization of the gains is as follows:

For institutions: $U_I^{OA} = P^{OA} + A^{OA} + I^{OA}(A^{OA}, N) - C^{OA}(APC)$.

For publishers: $U_\lambda^{OA} = \sum P_{OAj} * Pe_{OAj} - (FC_j + VC_j(P_{OA})) + I^{OA}(A^{OA}, N) = \pi_{\lambda_{OA}} + I^{OA}(A^{OA}, N)$.

- ***Both choose an Hybrid model***

The formalization of the gains is as follows:

For institutions: $U_I^H = [\lambda P^{OA} + (1-\lambda)P^C] + [\lambda A^{OA} + (1-\lambda)A^C] + I^H(A^H, N) - C^H(\sum \varphi * APC + S)$
$= \lambda[P^{OA} + A^{OA}] + (1-\lambda)[P^C + A^C] + I^H(A^H, N) - C^H(\sum \varphi * APC + S)$.

For publishers: $U_\lambda^H = \sum P_{Hj} * Pe_{Hj} - (FC_j + VC_j(P_H)) + I^H(A^H, N) = \pi_{\lambda_H} + I^H(A^H, N)$.

- ***Both choose an Subscription-based model***

The formalization of the gains is as follows:

For institutions: $U_I^C = P^C + A^C + I^C(A^C, N) - C^C(S)$

- For publishers: $U_\lambda^C = \sum P_{Cj} * Pe_{Cj} - (FC_j + VC_j(P_C)) + I^C(A^C, N) = \pi_{\lambda_C} + I^C(A^C, N)$

To simplify, we note:

$$\text{Institutions' payoffs per model} \begin{cases} U_I^{OA} = \alpha \\ U_I^C = \beta \\ U_I^H = \omega \end{cases}$$

$$\text{Publishers' payoffs per model} \begin{cases} U_\lambda^{OA} = \alpha' \\ U_\lambda^C = \beta' \\ U_\lambda^H = \omega' \end{cases}$$

Once the payoffs per model are determined, an order of preference of the institutions should be established according to the relationships described in the previous section. This this will allow to determine the institutions order of preferences of the different publishing strategies. Thus, we have shown the following relationships based on the literature:

- $A^{OA} > A^H > A^C$
- $P^{OA} > P^H > P^C$
- $I^H(A^H, N) \approx I^{OA}(A^{OA}, N) > I^C(A^C, N)$
- $C^H(\sum \varphi * APC + S) > C^C(S) > C^{OA}(APC)$

We can deduce that the OA publication strategy weakly dominates the other two strategies. Regarding the relationship between a subscription-based or hybrid model, we note that the hybrid publication model allows better research dissipation, better access to publication and a better impact than the subscription-based model. On the other hand, it costs much more and is not recommended by institutions and donors. Therefore, the cost constraint is so high that in the medium term, in the absence of a deal, institutions would converge to the subscription model to avoid paying twice for AO publications in the hybrid model.

In short, the preference relationship between the different strategies is as follows:

$$\boxed{U_I^{OA} > U_I^C > U_I^H \leftrightarrow \alpha > \beta > \omega}$$

Insofar as the costs borne by institutions and funders constitute the revenues of publishers, it can be established using the same reasoning that, on the one hand, $Pe_{Hj} > Pe_{Cj} > Pe_{OAj}$, and on the other hand $P_{OAj} > P_{Hj} > P_{Cj}$. Likewise, $I^H(A^H, N) \approx I^{OA}(A^{OA}, N) > I^C(A^C, N)$.

Since the cost / benefit ratio is the best in a hybrid model where institutions support both an OA and subscription-based model, publishers will prefer to stay on a hybrid model (dominant strategy). Moreover, symmetrically to the reasoning developed in the case of institutions, the bargaining power of publishers would become weaker in the case of an OA model, as would profits. This is partly explained by the fact that in a "read and publish" agreement, institutions will maximize their publication rate without necessarily increasing the earnings of publishers (or even the opposite). It can therefore be reasonably assumed that publishers would rather keep the subscription model than transition to an OA model.

In short, we can note:

$$\boxed{U_\lambda^H > U_\lambda^C > U_\lambda^{OA} \leftrightarrow \omega' > \beta' > \alpha'}$$

**The two players choose different strategies**
We can distinguish six situations:

- *Institutions choose an OA model and publishers a subscription-based model*

This scenario represents a situation in which the two players did not find an arrangement and neither side wants to compromise. In this case, the payoffs depend on their respective bargaining power and on the degree of integration of scientific publishing within institutions. This scenario may concern one institution facing a publisher, as it may concern also a group of institutions and a group of publishers who share the same strategy.

It is important to remember here that the bargaining power of institutions and funders is gradually increasing for several reasons:

- Institutions and research organizations produce scientific publications (source of income for publishers).
- Researchers (usually free of charge) provide peer review.
- Institutional and thematic open repositories have experienced considerable growth in recent years [74–77].
- The scientific community is becoming more and more organized and could integrate the evaluation of its own production (open peer review) [78–80].
- Open access publication becomes one of the research evaluation criteria [81–84]. Thus, researchers have less and less pressure to publish in traditional high-impact journals.

In this case, we can assume that institutions would substitute publishers who refuse to move to a fully OA model with open journals (and archives) on the one hand, and develop their own publication interfaces with an internal peer review process on the other. Therefore, the publisher's payoff in the case of a "subscription-based" strategy is zero.

Although the bargaining power of the institutions is high, they have an interest in reaching an amicable agreement with the publishers. Thus, we can reasonably assume that in the event of no agreement, the institutions will have to put in place mechanisms to control the quality of publications and ensure their sustainability. Considering the fact that the institutions are scattered across the world, such an organization would be very expensive for risky results. To be able to do without publishers, institutions must precisely determine the evaluation standards by community; have an equitable sharing of evaluation costs, and above all a commitment on the part of the various operators to participate in production and research evaluation.

From the point of view of the institutions, there are two major risks in the event of no agreement with the publishers:

- The degradation of the knowledge creation process due to the proliferation of behaviors of "invisible college" in the research evaluation [85]. Thus, researchers from the same scientific network can validate each other's publications without a real peer review. Or on the contrary, the increased risk of reprisals and settling of scores between researchers [86].
- The birth of a stowaway behavior, with "predatory" institutions that would benefit from the evaluation and publication process without any contribution.

Consequently, in terms of gains, in the event of no agreement, the institutions would have a lower gain ($\alpha^*$) than what they would obtain in the event of an agreement with the publishers. We notice:

$$\beta < \omega < \alpha^* < \alpha$$

- *Institutions choose an OA model and publishers a hybrid model*

In the event that institutions choose to publish only in OA, while publishers opt for a hybrid model, institutions will have the same gain as before ($\alpha^*$) as the majority of their publications will be fully open journals / archives. As the prices of APCs are relatively higher in Hybrids [87], the share of publications in these journals will be quite low. Thus, on the side of publishers, if institutions massively boycott publications in traditional journals, the gain would be much lower. We notice:

$$\omega'' < \alpha' < \beta' < \omega'$$

- *Institutions choose a subscription-based model and publishers an OA model*

This is a quite unlikely scenario given the recent evolutions in the publishing market in favor of free access (especially from institutions, governments and funders). Nevertheless, let us do the reasoning anyway. This situation may concern isolated institutions rather than a group of institutions or the market as a whole. This may be the case with an institution whose subscriptions to which researchers need are not expensive (or not numerous). The same goes for an institution whose interest in open access is moderate.

In this case, if the publishers to which they are subscribed decide to publish everything in open access without any substantial change in the price, their payoff will be the same in the case of an "amicable" agreement or a deal between the two parties. In other words, institutions will in this case benefit from all the benefits of OA (especially production and impact). The payoffs are therefore $\alpha$ for the institution. For publishers, the economic gains would be equal in the short / medium term to those of the subscription model, while the gains related to visibility and impact would be higher. We therefore define $\beta''$ such that:

$$\beta'' > \beta'$$

- *Institutions choose a subscription-based model and publishers a hybrid model*

For the same reason mentioned in the previous scenario, this situation is also unlikely. The difference with the previous scenario is that here publishers choose to make part of the publications freely accessible to increase their visibility. In this case, institutions would have the same advantages of a hybrid model at a lower cost (since the decision to make certain publications open access in a sustainable way comes from the publishers). Publishers get similar payoffs to those of the subscription-based model from an economic point of view but they get better visibility.

For institutions, we define the payoff $\beta^*$ with:
$$\beta < \beta^* < \alpha$$
For publishers, we define the payoff $\beta'''$ with:
$$\alpha' < \beta' < \beta''' < \beta'' < \omega'$$

- *Institutions choose a hybrid model and publishers an OA model*

This situation can rather concern an institution vis-à-vis a publisher than the entire market; it would suggest that the institution is willing to pay more to benefit from the notoriety of hybrid journals (the search for a high impact is greater than that of openness, for reasons of international rankings for example). If the publisher decides to publish everything in open access, the payoffs for the institutions will be that of the hybrid model, insofar as the publisher is rational (see axiom 4). Therefore, it will maintain a high price equivalent to that of the hybrid model (even if it publishes everything in open access).

For institution, we define the payoff $\omega^*$ with:
$$\omega < \omega^* < \beta < \alpha$$
$$\alpha^* \approx \omega^*$$

The publisher's gain is that of the hybrid model: $\omega'$.

- *Institutions choose a hybrid model and publishers a subscription-based model*

This scenario corresponds to a situation of no agreement between the two players. Thus, if publishers choose a completely "closed" model, institutions will massively boycott closed journals towards open journals or open archives / interfaces (herd behavior will then take place - see axiom 3). In this case, the closed publishers' payoff will be zero, while the institutions payoff will be equal to (see the first scenario described in the case of two different strategies) :
$$\alpha^* < \alpha$$

# Equilibria of the game

In this section, we represent the model in a strategic form to analyze the equilibria according to different situations presented above. To do this, let us recall here the order of preference of the gains according to the strategies:

- For institutions: $\alpha > \alpha^* \approx \omega^* > \beta^* > \beta > \omega$
- For publishers: $\omega' > \beta''' > \beta'' > \beta' > \alpha'' > \alpha' > \omega''$

Each player has three strategies: OA, Hybrid (H) and Subscription-based (C). We study the equilibria in pure strategies and in mixed strategies. We use the concepts of strategic dominance and Nash equilibrium (best response) to determine the equilibria.

Table 4 summarizes the payoffs according to publishers and institutions' strategies:

**Table 4: General payoff matrix**

|  |  | Publisher | | |
|---|---|---|---|---|
|  |  | OA | C | H |
| Institution | OA | $(\alpha, \alpha')$ | $(\alpha^*, 0)$ | $(\alpha^*, \omega'')$ |
|  | C | $(\alpha, \beta'')$ | $(\beta, \beta')$ | $(\beta^*, \beta''')$ |
|  | H | $(\omega^*, \omega')$ | $(\alpha^*, 0)$ | $(\omega, \omega')$ |

**Equilibrium in pure strategies**

## Equilibrium by iterative elimination of dominated strategies

Let us now seek equilibrium by iterative eliminating dominated strategies (IEDS). By definition, a dominated strategy (weakly or strictly) is a strategy that a player never chooses regardless of the choice of other players.

Thus, we can establish the following relationships between the different strategies (tables 5a and 5b):

**Table 5a: Institution's payoff according to the publisher's strategy**

|  | publisher's strategy | | | | | |
|---|---|---|---|---|---|---|
|  | OA |  | C |  | H | |
| Institution's payoffs | $\alpha$ | > | $\alpha^*$ | = | $\alpha^*$ | |
|  | $\alpha$ | > | $\beta$ | < | $\beta^*$ | |
|  | $\omega^*$ | ≈ | $\alpha^*$ | > | $\omega$ | |

**Table 5b: publisher's payoff according to the institution's strategy**

|  | Institution's strategy | | |
|---|---|---|---|
|  | OA | C | H |
| publisher's payoffs | α' < | β'' < | ω' |
|  | 0 < | β' > | 0 |
|  | ω'' < | β''' < | ω' |

For institutions, OA strategy weakly dominates strategy C and strictly dominates strategy H. For publishers, OA strategy is strictly dominated by strategy C and weakly dominated by strategy H.

Tables 6a and 6b show the process of iterative elimination of dominated strategies from institutions and publishers respectively.

**Table 6a: Equilibrium using iterative elimination of dominated strategies of institutions**

|  |  | Publisher | | |
|---|---|---|---|---|
|  |  | OA | C | H |
| Institution | OA | (α, α') | (α*, 0) | (α*, ω'') |
|  | C | (α, β'') | (β, β') | (β*, β''') |
|  | H | (ω*, ω') | (α*, 0) | (ω, ω') |

Since the OA strategy dominates the other two, institutions will choose it every time, whatever the choice of publishers. In this case, we eliminate lines C and H for the institution. On the publisher's side, there remain the three strategies OA, C and H with payoffs of, $\alpha'$, 0 and $\omega''$ respectively. As $\alpha' > \omega'' > 0$, the publisher will in this case choose the OA strategy as well. (OA, OA) is therefore the equilibrium if we start iterative elimination from institutions.

**Table 6b: Equilibrium using iterative elimination of dominated strategies of publishers**

|  |  | Publisher | | |
|---|---|---|---|---|
|  |  | OA | C | H |
| Institution | OA | (α, α') | (α*, 0) | (α*, ω'') |
|  | C | (α, β'') | (β, β') | (β*, β''') |
|  | H | (ω*, ω') | (α*, 0) | (ω, ω') |

If we start iterative elimination from publishers, the equilibrium obtained is different. Thus, the OA strategy is strictly dominated by the two strategies C and H in the case of publishers. By eliminating it, there will remain for the latter two strategies C and H, which becomes strictly dominant. Therefore, in the next iteration we can also eliminate strategy C for publishers. By iterative elimination of dominated strategies, the only strategy left to publishers is H. Institutions must therefore choose the strategy (s) with the highest payoff. As $\alpha^* > \beta^* > \omega$, the institution will in this case choose the OA strategy as well. (OA, H) is therefore the equilibrium if we start iterative elimination from publisher.

## Nash equilibrium according to the principle of best response

In this section, we use the best response principle to determine the Nash equilibrium. The equilibria determined with this method are by definition stable. In Nash equilibrium, each player chooses the best response (strategy) given the other player's choice. We speak of a Nash equilibrium if both players simultaneously choose the same strategy.

**Table 7: Nash equilibrium (using mechanism of best response)**

|  |  | Publisher |  |  |
|---|---|---|---|---|
|  |  | OA | C | H |
| Institution | OA | 🔵($\alpha$, $\alpha'$)🟢 | 🔵($\alpha^*$, 0) | 🔵($\alpha^*$, $\omega''$) |
|  | C | 🔵($\alpha$, $\beta''$)🟢 | ($\beta$, $\beta'$) | ($\beta^*$, $\beta'''$) |
|  | H | ($\omega^*$, $\omega'$)🟢 | 🔵($\alpha^*$, 0) | ($\omega$, $\omega'$)🟢 |

For example, if an institution chooses an OA model, the publisher can choose either an OA model and have a payoff $\alpha$, or a subscription-based model and have zero payoff, or a hybrid model and have a payoff $\omega''$. As $\alpha > \omega'' > 0$, the best answer in this case for the publisher is to choose an OA strategy. We apply the same reasoning with the two strategies (C and H) and we obtain the best answers from the editors presented in green in table 7.

On the institutions side, if the publishers choose an OA model, the best response in this case is to choose an OA or subscription-based model, since their payoffs are identical. Likewise, we apply the same reasoning and we obtain the payoffs presented in blue in table 7.

As we can see, there are two Nash equilibria: (OA, OA) and (C, OA). The first consists of mutual convergence towards an open access model, while the second consists of a paradigm shift on the part of publishers to move towards an OA model without affecting the cost of subscriptions.

**Mixed- strategies equilibrium**

Let us now analyze the equilibrium of this game in mixed strategies. In game theory, mixed strategy analysis involves assigning a probability to each of the players' strategies. This provides a more general basis for analysis and allows the study of equilibrium in an uncertain environment [88].

Since OA and H strategies the most likely strategies for institutions and publishers (as shown in the equilibria obtained by IEDS), we analyze the equilibrium with only these two strategies. Choosing a subscription-based model is unlikely in the current scientific context. Thus, institutions will choose an "OA" strategy with probability p, and an "H" strategy with probability (1-p). Symmetrically, publishers will choose an "OA" strategy with probability q and an "H" strategy with probability (1-q). The payoff matrix can therefore be represented as follows (table 8):

**Table 8: General payoff matrix (in mixed strategies)**

|  |  | Publisher | |
|---|---|---|---|
|  |  | OA (**q**) | H (**1-q**) |
| Institution | OA (**p**) | $(\alpha, \alpha')$ | $(\alpha^*, \omega'')$ |
|  | H (**1-p**) | $(\omega^*, \omega')$ | $(\omega, \omega')$ |

The expected gains are obtained as follows:

- For institutions:

$$q\alpha + \alpha^*(1-q) = q\omega^* + \omega'(1-q)$$

$$q = \frac{(\alpha - \alpha^* - \omega^* + \omega')}{\omega' + \alpha^*}$$

$$Best\ response\ for\ institutions \begin{cases} if\ q = \frac{(\alpha - \alpha^* - \omega^* + \omega')}{\omega' + \alpha^*} \to p \in [0,1] \\ if\ q > \frac{(\alpha - \alpha^* - \omega^* + \omega')}{\omega' + \alpha^*} \to p = 1 \\ if\ q < \frac{(\alpha - \alpha^* - \omega^* + \omega')}{\omega' + \alpha^*} \to p = 0 \end{cases}$$

This means that publishers will prefer an OA strategy for $q > \frac{(\alpha - \alpha^* - \omega^* + \omega')}{\omega' + \alpha^*}$. In other words, if the probability that institutions choose an OA model exceeds the ratio $p$, publishers would benefit from choosing the same strategy ($q = 1$). Whereas if the value $q < \frac{(\alpha - \alpha^* - \omega^* + \omega')}{\omega' + \alpha^*}$, publishers would benefit from choosing the hybrid strategy ($q = 0$).

- For publishers:

$$p\alpha' + \omega'(1-q) = p\omega'' + \omega'(1-q)$$

$$p\alpha' = p\omega''$$

$$\alpha' = \omega''$$

$$Best\ response\ for\ publishers \begin{cases} if\ \alpha' = \omega'' \to p \in [0,1] \\ if\ \alpha' > \omega'' \to p = 1 \\ if\ \alpha' < \omega'' \to p = 0 \end{cases}$$

On the institutional side, the best answer depends on the order of preference between $\alpha''$ and $\omega$. As shown above: $\alpha'' > \omega$. Therefore, the best response from institutions is to play OA regardless of the choice of publishers. In other words, $p = 1$ for $q \in [0,1]$.

Graphically, we can represent the best responses as follows:

**Fig. 4. Graphical representation of the best responses in mixed strategies**

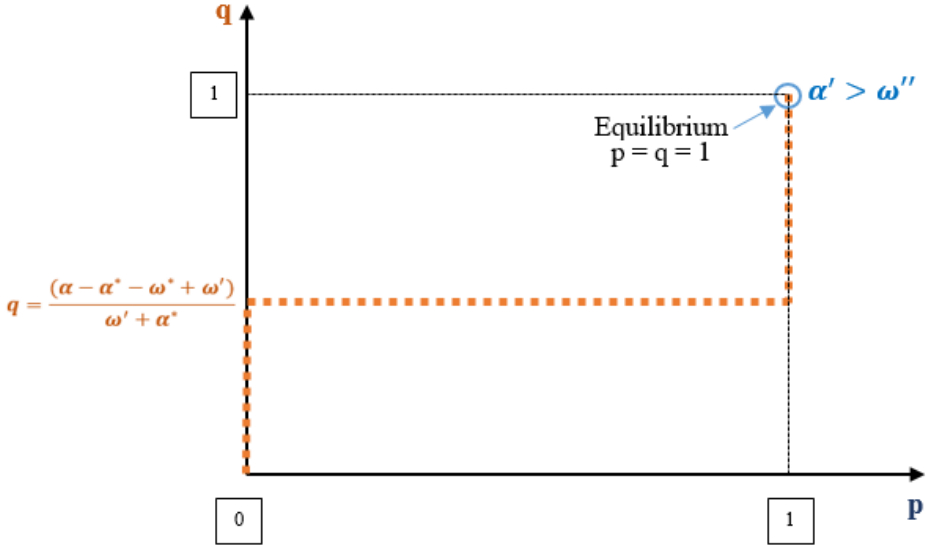

In mixed strategies, equilibrium is obtained when both players choose a strategy with the same probability. As can be seen in figure 4, the two best responses in terms of probabilities intersect in p = q = 1. This therefore corresponds to a combination of strategies (OA; OA). Therefore, the pure strategy Nash equilibrium of choosing an OA model for both players is also the only mixed strategy Nash equilibrium.

**Summary of equilibria according to the method used**

Table 9 summarizes the different equilibria obtained in pure strategies and in mixed strategies.

**Table 9: Equilibria in pure and mixed strategies**

| Type | Method | Equilibrium | Comment |
|---|---|---|---|
| Pure strategies | IEDS | (OA , OA) | IEDS of institutions leads to equilibrium (OA, OA) |
| | | (OA , H) | IEDS of publishers leads to two equilibria (C, H) and (H, H) |
| | Best response (Nash equilibrium) | (OA , OA), (C, OA) | The best-response equilibrium for the publisher is to choose an OA strategy regardless of the choice of institutions. For institutions, if publishers choose to publish everything in OA (without change of contract), staying in a subscription-based system is also a good strategy (this situation is unlikely). |
| Mixed strategies | Probabilities | (OA , OA) | In an uncertain environment, there is one and only one equilibrium, which is the publication in OA. |

## Generalization and convergence assumption

The strategic choices of the actors (institutions, groups of institutions, etc.) are rather made at the individual level. In the market, the three publishing models coexist in the short / medium term, with different proportions (market shares). The rise of the open access movement has turned scientific publishing market upside down in favor of hybrid and OA models. As shown previously, the publication in OA represents a Nash equilibrium in both pure strategy and mixed strategy. Thus, sometimes out of rationality and sometimes out of mimicry, institutions (at the instigation of researchers) are increasingly opting for OA publication. Aware of this movement, publishers have adapted by widely adopting the hybrid model.

A keen observer can easily notice 3 periods that the scientific publishing market has passed through, shown in the figures below. Each curve represents the share of reviews according to each of the three economic models. For any given year, the sum of the three curves is equal to 100%.

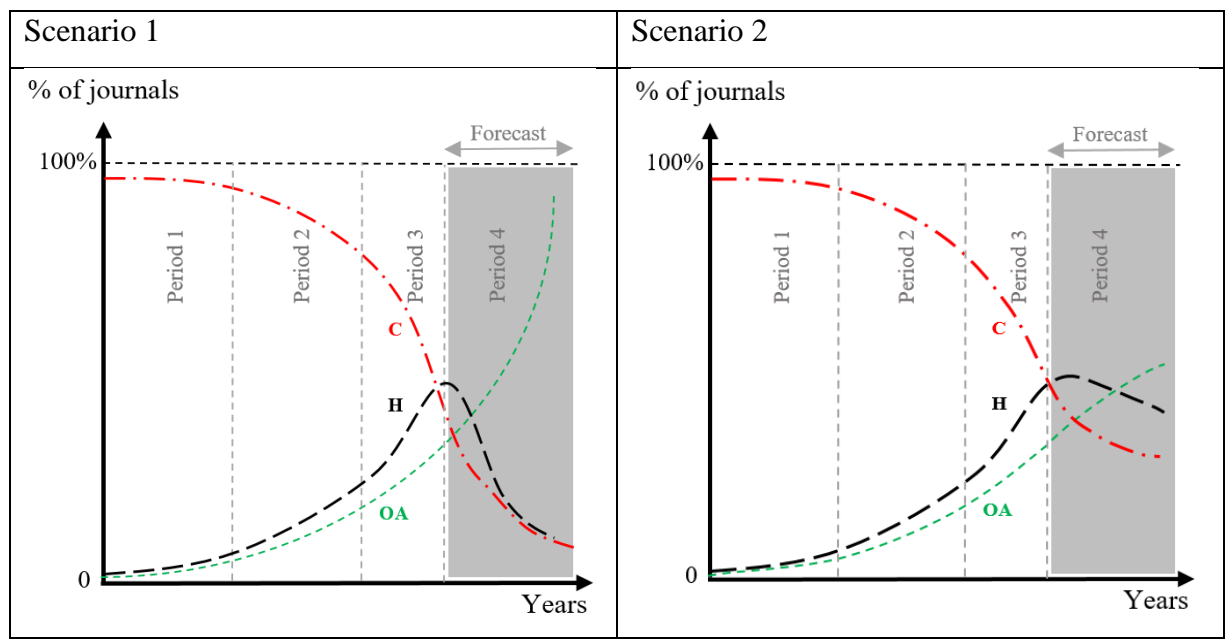

- Period 1 corresponds to the mid-1990s - early 2000s. With the democratization of the Internet, open access has grown in popularity. Thus, several open access journals were created, but also several open archives and indexes like arXiv.org (the first free scientific archive) and Medline.

- During period 2, several initiatives from the scientific community and institutions have emerged. This period spanned from the mid-2000s to the mid-2010s, it marks the rise of the open science movement. During this period, we witnessed a form of institutionalization of open access publishing with the development of new economic models. This is for example the case with the for-profit open access publisher BioMed Central launched in 2000, or the PLOS journals. This period also marks the proliferation of the hybrid model, particularly from 2013.
- Period 3 is a period of political awareness of the interest of the OA (from the end of 2014 to the present day). This awareness goes far beyond the scientific communities to reach university governors, politicians, funders, and the media. OA is increasingly becoming one of the criteria for evaluating research (eg in the United Kingdom). Several national open science plans have emerged all over Europe and in high-income countries. Likewise, several missions have been launched to reflect on the substance of the contracts with the major publishers.
- Period 4 is representative of the future of the world of scholarly publishing, taking into account the latest developments. There are two scenarios: the first one is more "optimistic" in which the transition to an OA model will be done quickly. In the second scenario, the transition will be slower with greater resistance from hybrid and subscription-based models. In any case, the future development of the publishing market depends on several factors: the bargaining power of publishers, open science policies, the strength of coalitions of scientific communities and the pressure they can exert, etc.

The game theory analysis presented in the previous section showed that the OA model leads to a stable equilibrium (Nash equilibrium) in both pure and mixed strategy. If we apply here the rationality hypothesis and that of memetic behavior, we can easily deduce that the OA model will inexorably be the dominant model on the entire market in the medium term: All the players will converge on an OA publication model.

## Conclusion and discussion

Through this paper, we sought to model objectively the scientific publishing market as presented in the literature and following public and institutional policy developments in open science. Three main economic models currently coexist in the scholarly publishing market: the traditional model (subscription-based), the open access model (author-pays) and the hybrid model (which combines the two). Since it is the institutions that support the publication costs and define the editorial strategies, the choice of publication of the authors depends greatly on that of their institution. It is for this reason that the strategic reasoning has been constructed in this article from institutions. Institutions therefore make trade-offs based on the choices offered by publishers. In order to determine the preferred model for each of the two players (institutions and publishers), we studied the payoffs they would obtain in each economic model.

We have assumed that the payoffs depend on three things: production and access, academic and societal impact, and finally costs (for institutions) and profits (for publishers). Using the results obtained in the literature we have shown that the model that maximizes the utility of institutions is that of open access. Thus, it allows a better dissemination of research results, and increase the productivity of researchers. This model also allows for better visibility and a relatively high impact. On this last point, it is the hybrid model, which appears the best insofar as many well-established journals on the market are of the hybrid type and therefore receive articles of better quality and obtain a high impact. On the other hand, the hybrid model is the most expensive of all and it is less interesting than the OA model in terms of production and access to publications. On the side of the editors, it is the hybrid model, which maximizes the gains. It is therefore preferred that the subscription-based model, which in turn is preferred over the OA model.

We have shown that there is a Nash equilibrium in which institutions and publishers converge towards an OA model. There is also another Nash equilibrium in which OA publishing is publisher-led (which is unlikely). These equilibria are Pareto optimal, because they results using best response method.

When we proceed by eliminating the dominated strategies of publishers, the equilibrium obtained is as follows: OA model for institutions and hybrid for publishers. On the opposite, when we proceed by eliminating the dominated strategies of the institutions, the equilibrium obtained is the OA model for both players. The direction of elimination of dominated strategies depends greatly on the bargaining power of the two players. The bargaining power of institutions has increased dramatically in recent years with the breakthrough of the open science movement. Thus, in addition to the fact that it is the institutions that produce and evaluate scientific publications, political attention is increasingly turned to open science, which has become one of the evaluation criteria for researchers. This puts a lot of pressure on publishers who are starting to make some compromises. The hybrid model therefore appears to be unstable in the short term. On the institutional side, it is also difficult and risky to do without publishers, particularly from the point of view of sustainability and quality control of publications. Thus, several perverse effects and abuses can be accentuated if the institutions alone ensure the entire publication process. In particular, the degradation of the evaluation because of the effects of networks and "invisible colleges" on the one hand, and of stowaways on the other.

Institutions and publishers therefore have an interest in converging and reaching an agreement, even if it means making some compromises: the "damage" of a no-deal is so high that the two will end up converging on an open-access deal. This is true for both players. Thus, if institutions do not compromise would jeopardize the process of knowledge creation and research assessment. Likewise, if publishers choose to remain in old (subscription-based) or transitional (hybrid) economic models risk a broad boycott of the scientific communities and a drastic drop in their profits.

The analysis of equilibrium in mixed strategies confirms this result. The integration of probabilities in the choice of strategies shows that publishers would converge towards an open access model if and only if the probability that institutions choose to publish in open access exceeds a certain threshold. We then showed that the probability of choosing the open access model is always equal to one regardless of the choice of the publishers. This means that there is only one equilibrium in mixed strategy, which is the choice of the open access model for institutions and publishers.

By analyzing market forces and the latest policy developments in research evaluation and funding, this article shows that the world of science is inevitably moving towards an agreement around open access. The only variable unknown to date is the speed of convergence. This speed of convergence is linked to several factors such as the importance given by funders to open science and the pressure exerted by researchers and scientific communities. While open access publication is starting to be integrated into the evaluation criteria in several countries such as the United Kingdom and France, there is still a long way to go before the journals that are born in open access establish themselves on the market and change the publication practices of researchers. On these questions, there are scientific communities more advanced than others and the transition to a fully open access model could be done step by step and community by community. For the most advanced of them, the transition seems to be close, for others the path is well traced.